\documentstyle [prl,multicol,aps,epsfig]{revtex}
\begin{document}
\title{Theory of spin relaxation in magnetic resonance force microscopy}
\author{D. Mozyrsky$^1$, I. Martin$^1$, D. Pelekhov$^2$, and P. C. Hammel$^2$}
\address{$^1$Theoretical Division, Los Alamos National Laboratory, Los Alamos, NM 87545, USA\\
$^2 $Department of Physics, Ohio State University, Columbus, OH 43210, USA}
\date{Printed \today}
\maketitle
\begin{abstract}
We study relaxation of a spin in magnetic resonance force microscopy (MRFM)
experiments.  We evaluate the relaxation rate for the spin caused by
high-frequency mechanical noise of the cantilever under the conditions of
adiabatic spin inversion. We find qualitative agreement between the obtained
relaxation time and the experimental results of Stipe {\it et al.} [Phys. Rev.
Lett. 87, 277602 (2001)]. Based on our analysis, we propose a method for
improving the MRFM sensitivity by engineering cantilevers with reduced tip
positional fluctuations.

\end{abstract}
\pacs{PACS Numbers: XXXXX}
\begin{multicols}{2}

Magnetic resonance force microscopy (MRFM) has proved to be a powerful tool in
studying magnetic properties of materials~\cite{1-1,1-2,1-3,2}. Unlike
conventional magnetic resonance techniques, MRFM allows one to probe
magnetization with a nanometer-scale  spatial resolution, which is important
for practical realization of spintronic~\cite{3} and quantum information
processing devices~\cite{4}. Measurement of magnetizations produced by several
hundreds of electronic magnetic moments has recently been reported~\cite{2}.
Attempting to reach the single spin resolution, MRFM faces a number of
experimental challenges. The observed reduction of the cantilever's quality
factors for small separations between the tip and the sample~\cite{5-1,5-2} and
fast spin relaxation~\cite{2} pose a problem in resolving a single spin signal.
In this paper we analyze the latter effect. In particular, we study relaxation
of a spin in the resonance slice~\cite{2} due to the thermal vibrations of the
cantilever. We evaluate the spin relaxation rate and find a reasonable
agreement with rates observed experimentally. We also propose that by shape
engineering of the cantilever it is possible to filter the high frequency noise
thus reducing the spin relaxation.

\begin{figure}
\includegraphics[angle=90, width = 3.3 in, height = 2.2 in]{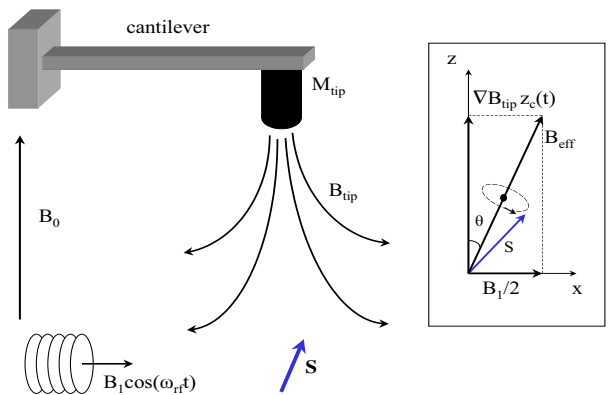}
\caption{Model setup for MRFM.}
\end{figure}

In this work we consider the experimental setup used in Ref.~\onlinecite{2}.
One end of the cantilever is fixed, while the other end, to which micron size
magnetic particle is attached, is driven to oscillate in a direction normal to
the surface of the sample; see Fig.~1. The sample is doped with spin $1/2$
magnetic impurities. The smallest flexural-mode eigenfrequency of the
cantilever will be denoted by $\omega_0$ and the amplitude of the oscillations
of the tip by
$z_0$. In typical MRFM experiments $\omega_0$ is several {\rm kHz}  and
$z_0\sim 0.1$-30~{\rm nm}.  The motion of the
cantilever's magnetic tip produces an oscillating magnetic field ${\bf B}_{\rm
tip}({\bf r},t)$ at the spatial position ${\bf r}$ at time $t$ (the origin is
taken as the equilibrium position of the tip). There is an additional constant
magnetic field ${\bf B_0}$ applied normally to the sample, and a linearly
polarized microwave magnetic field
${\bf B}_1 \cos{(\omega_{\rm rf}t)}$ in a direction perpendicular to ${\bf B_0}$.

The magnetic resonance condition is achieved as follows. Let us assume for
simplicity that a single impurity spin is positioned at point ${\bf r}$
immediately below the cantilever tip, as shown in Fig.~1. Then, the magnitude
of the field produced by the magnetic tip at the spin's position is ${B}_{\rm
tip}({\bf r},t) = {B}_{\rm tip}^0({\bf r}) + \nabla_z {B}_{\rm tip}^0({\bf r})
z(t)$, where $B_{\rm tip}^0({\bf r})$ is the field of the tip in equilibrium,
$z(t)$ is cantilever's displacement, $|{\bf r}| \gg z(t)$. Let the frequency
$\omega_{\rm rf}$ of the microwave field be chosen such that $ g\mu_B|B_{\rm
tip}^0({\bf r}) - B_0| \simeq \hbar\omega_{\rm rf}$, where $g$ is electronic
$g$-factor and $\mu_B$ is Bohr magneton.  Then, in the {\it rotating reference
frame} that rotates with frequency $\omega_{\rm rf}$ around the normal to the
surface~\cite{Slichter}, the spin experiences an effective magnetic field (see
the inset to Fig.~1)
\begin{equation}
{\bf B}_{\rm eff} = \nabla_z {\bf B}_{\rm tip}^0({\bf r}) z (t) + {\bf B}_1/2\, .\label{a1}
\end{equation}
\noindent As a result, the spin precesses around this effective field, thus following the
direction of ${\bf B}_{\rm eff}$, as well as cantilever's displacement,
provided the latter varies with time adiabatically, i.e., $\omega_0 \ll
\omega_{\rm eff} = g \mu_B |{\bf B}_{\rm eff}|/\hbar$. Thereby, the spin
direction follows the cantilever motion and the spin exerts an oscillating
force on the cantilever's tip with the amplitude, $F=
\mu_B\nabla_z B_{\rm tip}^0$. In Ref.~\onlinecite{2} the shift of the
cantilever's frequency
$\Delta\omega_0$ due to the magnetization of the resonant spins has been
measured. The signal that corresponds to roughly
$100$ fully polarized spins was observed to decay on a time scale of $100$ {\rm ms}, thus
indicating that the induced magnetization of the resonant spins relaxes due to magnetic
fluctuations or ``noise'' whose origin we will discuss.

In this letter we show that the main source of the noise causing spin
relaxation is likely to be related to the cantilever's thermal vibration. The
cantilever displacement can be written as $z(t) = z_c(t) +
\delta z(t)$, where the first term, $z_c(t) = z_0\cos{(\omega_0 t)} + z_1$, is due to the regular (driven) oscillation
of the cantilever tip, while $\delta z(t)$ is due to the thermal motion of the
tip.  Here, $z_1$ defines the relative position of a spin within the resonant
slice ($|z_1| \leq z_0$). Substituting the above
$z(t)$ into Eq.~(\ref{a1}) we obtain an effective Hamiltonian for a spin in the
rotating frame
\begin{equation}
H = H_0(t) + n(t) s_z, \label{a2}
\end{equation}
\noindent where $H_0 = g\mu_B[\nabla_z B_{\rm tip}^0 z_c(t) \, s_z + (B_1/2) \,
s_x]$, ${\bf s}$ is electronic spin, and $n(t) =  g\mu_B\nabla_z B_{\rm tip}^0 \delta z(t)$. In
the following analysis we set $g=2$ and the Planck constant $\hbar=1$, unless stated otherwise. We
assume that the noise $n(t)$ is Gaussian with the correlation function $K(t_1-t_2) = \langle
n(t_1)n(t_2)\rangle$ to be specified below. We derive the Bloch equations for the spin from the
dynamics governed by the Hamiltonian~(\ref{a2}). We introduce the Keldysh contour and define a real
time partition function along the contour as
\begin{equation}
{\cal Z} = {\cal T}_c \exp{\left(-i\int_\infty^{-\infty} H dt\right)}
\exp{\left(-i\int_{-\infty}^\infty H dt\right)}, \label{a3}
\end{equation}
\noindent where ${\cal T}_c$ denotes ordering along the contour~\cite{Keldysh}. Thus points on the
forward branch $(-\infty \rightarrow \infty)$ are ordered with increasing times, while points on
the return branch $(\infty \rightarrow -\infty)$ are ordered with decreasing times. The
superscripts $f$ and $r$ on $t$ will indicate to which contour $t$ belongs. In Eq.~(\ref{a3}) the
time ordering operator ${\cal T}_c$ sets operators on the return branch in front of the operators
on the forward branch of the contour.  The expectation value of an operator $O(t)$ ($O = s_x, s_y,
s_z$) can be obtained by using the partition function ${\cal Z}$ as $\langle O(t) \rangle = {\rm
Tr}\left[ {\cal T}_c O(t^f) {\cal Z} \right]$, where the trace is taken over the states of the
spin and over the distribution of the classical noise $n(t)$.  Averaging the partition function
${\cal Z}$ over the noise under the assumption that the noise is Gaussian, we obtain
\begin{eqnarray}
 \langle{\cal Z}\rangle_n &=& {\cal T}_c \,e^{-i\int_\infty^{-\infty} H_0 dt}
 e^{-i\int_{-\infty}^\infty H_0 dt}\nonumber\\
 & \times & \exp{\left[-{1 \over 2} \int_{-\infty}^\infty K(t_1 - t_2) s_z^a (t_2) s_z^a (t_1) dt_1
 dt_2\right]}, \label{a5}
\end{eqnarray}
\noindent where $s_z^a(t)=s_z^f(t) - s_z^r(t)$. The effective action for the spin defined by
Eq.~(\ref{a5}) is generally non-local. It can be significantly simplified by introducing the
Bloch-Redfield approximation~\cite{Slichter}. If the interaction of the system with the noise is
weak, one can expect that on a scale of the noise correlation time, $\tau_c$, the spin's evolution
will be mostly determined by the unperturbed Hamiltonian $H_0$. That is, in the non-local part of
the action one can replace $s_z^a(t_2)$ by $\exp{[i\int_{t_1}^{t_2} H_0(\tau)d\tau]} s_z^a(t_1)
\exp{[-i\int_{t_1}^{t_2} H_0(\tau)d\tau]}$. The latter operator can be evaluated assuming that
$H_0(\tau)$ varies adiabatically. By substituting the resulting expression into Eq.~(\ref{a5}),
the non-local part of the effective action in Eq.~(\ref{a5}) can thus be approximated as
\begin{eqnarray}
\Delta {\cal S} = -{1 \over 2} \int \Big\{\cos{\theta}\left[\cos{\theta} s_z^a(t) + \sin{\theta}
s_x^a(t)\right]S_n[0]\nonumber\\
+ \sin{\theta}\left[\sin{\theta} s_z^a(t) - \cos{\theta} s_x^a(t)\right]S_n[\omega_{\rm
eff}(t)]\Big\}dt\, , \label{a6}
\end{eqnarray}
\noindent where $S_n[\omega] = \int_{-\infty}^\infty K(t)\exp{(i\omega t)} dt$ is the power
spectrum of the noise $n(t)$, $\omega_{\rm eff}(t) = g\mu_B\{[\nabla_z B_{\rm
tip}^0 z_c(t)]^2 + B_1^2/4\}^{1/2}$, and $\sin{\theta} = g\mu_B
B_1/2\omega_{\rm eff}(t)$. In the derivation of Eq.~(\ref{a6}) we have assumed
that $\omega_{\rm eff}/(d\omega_{\rm eff}/dt)\gg\tau_c$. The effective action
in Eq.~(\ref{a6}) is now local and allows for a straightforward derivation of
equations of motion for the spin components $s_{x,y,z}$. Using
${\dot s}_{x,y,z} = i[H_{\rm eff}\, ,s_{x,y,z}]$, where $H_{\rm eff}$ is defined by
Eqs.~(\ref{a5}) and (\ref{a6}), and ordering the operators according to Eq.~(\ref{a3}), we obtain a
set of Bloch equations for the magnetization components
\begin{mathletters}
\label{a7}
\begin{eqnarray}
&&{\dot s}_x = -\omega_{\rm eff}(t) \cos{\theta(t)} s_y + \beta(t) s_z - \alpha(t) s_x\,,\\
\label{a7a} &&{\dot s}_y = \omega_{\rm eff}(t) \cos{\theta(t)} s_x -\omega_{\rm eff}(t) \sin{\theta(t)}s_z
- \alpha(t) s_y\, ,\\
\label{a7b} &&{\dot s}_z = \omega_{\rm eff}(t) \sin{\theta(t)}s_y\, . \label{a4c}
\end{eqnarray}
\end{mathletters}
\noindent In Eqs.~(\ref{a7}), terms with coefficients $\alpha$ and $\beta$ describe
relaxation of the magnetization due to noise, $\alpha(t) =
\{\cos^2{\theta} S_n[0] +
\sin^2{\theta}S_n[\omega_{\rm eff}(t)]\}/2$, and $\beta(t)=\sin{2\theta}\{S_n[0]-S_n[\omega_{\rm
eff}(t)]\}/4$. By solving the Bloch equations in the adiabatic limit, i.e., by assuming that the
coefficients in Eqs.~(\ref{a7}) are slowly varying functions of $t$, we find that the component of
the  magnetization parallel to the effective field decays as
\begin{equation}
|s_{\rm eff}(t)| \sim \exp{\Big\{-{1 \over 2} \int_0^t \sin^2{\theta(\tau)}S_n[\omega_{\rm
eff}(\tau)]d\tau\Big\}}\, . \label{a8}
\end{equation}

Now we turn to the evaluation of the power spectrum of the magnetic noise produced by the
cantilever tip. The magnetic noise is related to the mechanical noise of the tip as $S_n[\omega] =
(g\mu_B\nabla_z B_{\rm tip}^0)^2 \int_{-\infty}^\infty dt \exp{(i\omega t)} \langle\delta z
(0)\delta z(t)\rangle$. It should be emphasized that we are interested in the mechanical noise of
the cantilever at frequencies very high (of order $\omega_{\rm eff}$) compared to the lowest
cantilever eigenfrequency $\omega_0$. As a consequence of the linearity of mechanical oscillator
the high frequency modes of the cantilever will be essentially unaffected by the driving force at
frequency $\omega_0$; rather it is the thermal driving force that is important.

The energy of the vibrating cantilever can be expressed as $E_c = \int_0^l dx [\rho (\partial_t
z)^2 + EI(\partial_x^2 z)^2]/2$, where $z(x,t)$ is the transverse displacement of the cantilever at
point $x$ and time $t$, $\rho$ is cantilever's linear mass-density, $E$ and $I$ are Young modulus
and moment of inertia of the cantilever's cross-section respectively, and $l$ is the cantilever's
length. The equation of motion for the free cantilever is $\rho \partial_{t}^2 z =
EI\partial_{x}^4z$~\cite{Landau}. The general solution to this equation can be written as $z(x,t) =
\sum_n z_n (t) \phi_n(x)$, with $\phi_n$'s satisfying the eigenvalue equations for $\omega_n$:
$\rho\omega_n^2\phi_n = EI\partial_{x}^4 \phi_n$, supplemented by appropriate boundary conditions
~\cite{Landau}. In the eigenfunction basis $\phi_n$, the energy of the cantilever can be expressed
in terms of the amplitudes $z_n$: $E_c = \sum_n [m {\dot z}^2_n + m\omega_n^2 z^2_n]/2$, where
$m=\rho l$ is cantilever mass. Hence, the cantilever can be modeled as a collection of independent
oscillators. In thermal equilibrium we then have that $\langle \delta z_n^\ast (\omega) \delta
z_{n^\prime}(\omega)\rangle = \delta_{n,n^\prime}(\pi k_B T /m\omega_n^2)[\delta(\omega_n +
\omega)+\delta(\omega_n - \omega)]$, where $k_B T$ is the temperature and the delta functions $\delta(\omega_n \pm
\omega)$ have effective width of order of damping coefficient for the $n$-th mode.
The magnetic noise power spectrum can be expressed as
\begin{equation}
S_n[\omega] = (g\mu_B\nabla_z B_{\rm tip}^0)^2 \sum\nolimits_n \phi_n^2(l) \langle |\delta
z_n(\omega)|^2\rangle\, ,\label{a9}
\end{equation}
\noindent where the eigenfunctions $\phi_n(x)$ are normalized, $\int_0^l \phi_n^2(x)dx = l$. The
power spectrum given by Eq.~(\ref{a9}) generally represents a set of distinct
peaks.  However the effective frequency of a spin,
$\omega_{\rm eff}(t)$, sweeps over a large number of peaks.  Moreover, the form of
Eq.~(\ref{a8}) indicates that it is the integral characteristics of the
spectrum that are relevant for the determination of the spin relaxation rate.
Therefore the sum in Eq.~(\ref{a9}) can be replaced by the integral, $\sum_n
\rightarrow \int (\partial n/\partial
\omega_n) d\omega_n$ and the finite width of the delta functions $\delta(\omega_n \pm \omega)$ can
be neglected. Then, taking into account that for $n \gg 1$, $\omega_n = (\pi n/l)^2
(EI/\rho)^{1/2}$, $\phi_n^2(l)=4$, and $\omega_0 = (3.52 / l^2) (EI/\rho)^{1/2}$, see
Ref.~\onlinecite{Landau}, Eq.~(\ref{a9}) yields the average noise spectral density
\begin{equation}
S_n^{\rm ave}[\omega] \simeq 7.5 {(g\mu_B\nabla_z B_{\rm tip}^0)^2 k_B T \over m {\omega_0}^{1/2}
\omega^{5/2}}\, ,\label{a10}
\end{equation}
Substitution of Eq.~(\ref{a10}) into Eq.~(\ref{a8}) gives the spin relaxation
rate. For large times, $t\gg\omega_0^{-1}$ and for $\nabla_z B_{\rm tip}^0
|z_0\pm z_1|\gg B_1$, the integral in Eq.~(\ref{a8}) can be readily evaluated
yielding
$|s_{\rm eff}(t)|\sim\exp{(-t/\tau_m)}$, where
\begin{equation}
{1 \over \tau_m} \simeq {3.4 \mu_B\nabla_z B_{\rm tip}^0\over\hbar}\left({k_B T
\over k_s \sqrt{z_0^2 - z_1^2}}\right)\left({\omega_0 \over \omega_1}\right)^{3/2}\,
.\label{a11}
\end{equation}
Here we have introduced spring constant $k_s = m\omega_0^2$ and
$\omega_1 = \mu_B B_1/\hbar$.

We can now compare the relaxation time given by Eq.~(\ref{a11}) with the
experimental results of Ref.~\cite{2}. For tip-sample separations of $800$ nm
the gradient of the magnetic field ($\nabla_z B_{\rm tip}^0$) was found to be
roughly $1 \times 10^5$ T/m. Taking other experimental parameters from
Ref.~\onlinecite{2}, we obtain from Eq.~(\ref{a11}) the maximum $\tau_m
\simeq 350$ ms (for $z_1 = 0$).  Spins located away from the center of the resonating
region will have shorter relaxation times.  This as well as the simplifying
theoretical assumptions about the cantilever geometry are likely to be
responsible for about five times difference between the theoretical estimate
for the longest relaxation time and the experimental results~\cite{2}.  We also
note that because of the relaxation time distribution, measurement of an
ensemble of spins is likely to yield a non-exponential MRFM signal decay.

In the mechanism discussed above, the spin relaxation is caused by the thermo-mechanical noise of
the high frequency modes of the cantilever tip. This suggests a way to reduce the relaxation rate
by engineering the shape of the cantilever to reduce this noise. The simplest approach that we
analyze here is to place a massive particle ($M$) on the tip.  This will effectively filter out
frequencies $\omega_n$ for which $n > m/M$.  For such high-frequency modes it is easy to show that
$\phi_n^2(l) \simeq 2\sqrt{\rho^3EI}/(M^2\omega_n)$.  For example, for a particle with $M = m/10$
and the same experimental parameters, the relaxation time would increase to $\tau \sim 5$ s.

In summary, we have found that high frequency thermo-mechanical noise
associated with high frequency modes of the cantilever can induce significant
spin relaxation in the magnetic resonance force microscopy, and is a likely
reason for the coherent signal loss in the recent high-sensitivity
experiments\cite{2}. To reduce the influence of this noise, we have proposed
cantilever shape engineering, which can lead to significant enhancement of MRFM
sensitivity.

The work was supported by the DARPA MOSAIC program. D.M. was
supported, in part, by the US NSF grants ECS-0102500 and
DMR-0121146.

\end{multicols}
\end{document}